\documentclass[aps,pra,twocolumn,groupedaddress]{revtex4} 

\usepackage{graphicx}
\usepackage{bm}
\usepackage{epstopdf}

\begin{document}


\title{Photoassociative Spectroscopy of a Halo Molecule in $^{86}$Sr}



\author{J. A. Aman, J. C. Hill, R. Ding, Kaden R. A. Hazzard, and T. C. Killian}
\affiliation{Rice University, Department of Physics and
Astronomy and Rice Center for Quantum Materials, Houston, Texas, 77251}
\author{W. Y. Kon}
\affiliation{Rice University, Department of Physics and
Astronomy and Rice Center for Quantum Materials, Houston, Texas, 77251}
\affiliation{School of Physical and Mathematical Sciences, Nanyang Technological University, Singapore}


\date{\today}

\begin{abstract}
We present two-photon photoassociation to the least-bound vibrational level of the X$^1\Sigma_g^+$ electronic ground state of the $^{86}$Sr$_2$ dimer and measure a binding energy of $E_b=-83.00(7)(20)$\,kHz. Because of the very small binding energy, this is a halo state corresponding to the scattering resonance for two $^{86}$Sr atoms at low temperature. The measured binding energy, combined with universal theory for a very weakly bound state on a potential that asymptotes to a van der Waals form, is used to determine an $s$-wave scattering length $a=810.6(12)$\,$a_0$, which is consistent with, but substantially more accurate than the previously determined $a=798(12)\,a_0$ found from mass-scaling and precision spectroscopy of other Sr isotopes. For the intermediate state, we use a bound level on the metastable $^1S_0-{^3P_1}$ potential. Large sensitivity of the dimer binding energy to light near-resonant with the bound-bound transition to the intermediate state suggests that $^{86}$Sr has great promise for manipulating atom interactions optically and probing naturally occurring Efimov states.

\end{abstract}


\maketitle



\section{Introduction\label{introduction}}

Weakly bound ground-state dimers are of great interest in ultracold atomic and molecular physics. In the extreme case of a scattering resonance, the least-bound state represents an example of a quantum halo system \cite{jrf04} with spatial extent well into the classically forbidden region. Halo molecules show universality, meaning that molecular properties such as size and binding energy can be parameterized by the single quantity, $s$-wave scattering length $a$, independent of other details of the atom-pair interaction \cite{kgj06,bha06}. For potentials that asymptote to a van-der-Waals form, an additional parameter, the van der Waals length $l_{\mathrm{vdW}}$, can be introduced for a more accurate description. Efimov trimers also exist in systems near a scattering resonance, influencing dimer and atomic scattering properties and introducing additional universal phenomena \cite{bha07,nen17}. Ultracold halo molecules are often associated with magnetic Feshbach resonances \cite{cgj10}, for which the scattering state and a bound molecular state can be brought near resonance by tuning a magnetic field.



\begin{figure}
  \includegraphics[width=3.3in,clip=true, trim=0 0 0 0, angle=0]{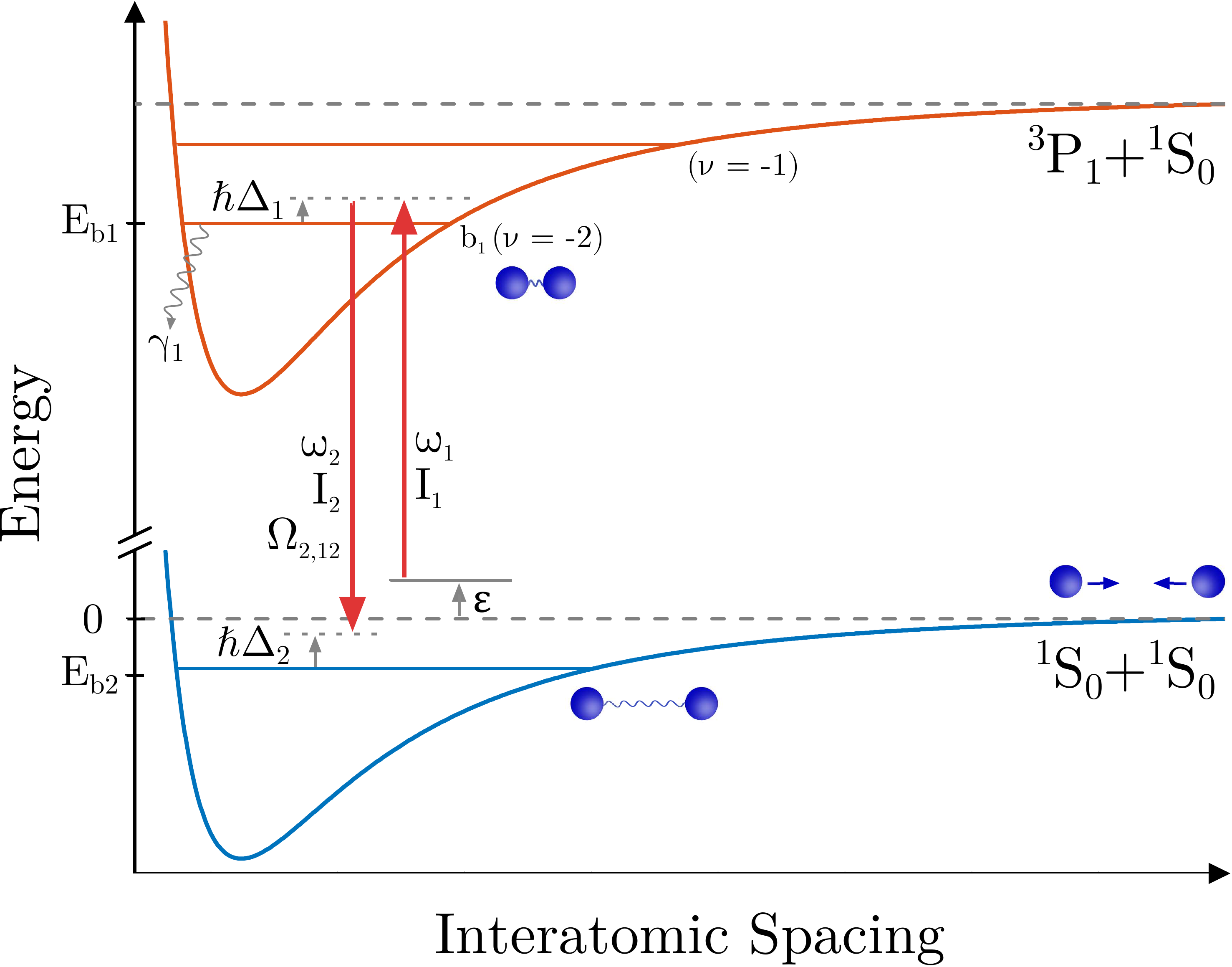}
  \caption{Two-photon photoassociation diagram. The energy of two well-separated $^1S_0$ atoms at rest is taken as zero. $\epsilon$ is the kinetic energy of the colliding atom pair. $E_{b1}$ is the unperturbed energy of the bound state of the excited molecular potential that is near resonance with the free-bound laser, which in these experiments is the second-least bound level of the excited molecular potential ($\nu=-2$). $E_{b2}$ ($<0$) is the unperturbed energy of the least bound state of the ground molecular potential. The photon of energy $\hbar \omega_1$ is detuned from $E_{b1}$ by $\hbar \Delta_1$, while the two-photon detuning from $E_{b2}$ is $\hbar \Delta_2$. The decay rate of $b_1$ is $\gamma_1$. Stark and collisional frequency shifts are neglected in this schematic.}
  
  \label{PASDiagram}
\end{figure}

Here we study the least-bound vibrational level of the X$^1\Sigma_g^+$ electronic ground state of the $^{86}$Sr$_2$ dimer (Fig.\ \ref{PASDiagram}), which is a naturally occurring halo molecule, meaning it exists in the absence of tuning with a magnetic Feshbach resonance. A well-known example of a naturally occurring halo molecule is the $^4$He$_2$ dimer \cite{lmk93,sto94,kgj06}. The least-bound vibrational level of the ground state of $^{40}$Ca$_2$, which was recently studied using similar methods \cite{pdt17}, is also very close to this regime.


There are important differences between halo molecules associated with magnetic Feshbach resonances and the naturally occurring halo molecule in $^{86}$Sr. With magnetic Feshbach resonances, the relevant scattering and bound molecular states lie on different molecular potentials, and single-photon magnetic-dipole transitions can be used to measure molecular binding energies with RF or microwave spectroscopy \cite{cgj10,cju05,thw05b}. Typically, this is done by first forming molecules through magneto-association and then driving bound-free or bound-bound transitions converting the halo molecule into a different state. Other methods include spectroscopy with an oscillating magnetic field \cite{thw05b}, a modulated optically controlled Feshbach resonance \cite{chx15}, and Ramsey-type measurements of atom-molecule oscillation frequencies \cite{ckt03}. It is also possible to efficiently populate halo states with a magnetic-field sweep \cite{grj03} or evaporative cooling \cite{jba03} near a magnetic Feshbach resonance \cite{cgj10}. These are powerful techniques for manipulating quantum gases of alkali metals and other open-shell atoms, for which there are many magnetic Feshbach resonances. Strontium, however, due to its closed-shell electronic structure, lacks magnetic Feshbach resonances in the electronic ground state.

In this work, we probe the halo state in $^{86}$Sr using two-photon Raman photoassociation (PA) \cite{jtl06}, in which two laser fields couple colliding atoms to the least-bound state of the ground molecular potential. We tune near resonance with an intermediate state that is bound in the $0_u$ potential  corresponding to the $^1S_0+ {^3P_1}$ asymptote at long range \cite{mmp08} (Fig.\ \ref{PASDiagram}). We accurately determine the $^{86}$Sr$_2$ binding energy, considering possible collisional frequency shifts and AC Stark shifts due to trapping and excitation lasers. Using the universal prediction for the binding energy, including corrections derived for a van der Waals potential \cite{gfl93,gao01,gao04}, we derive a more accurate value of the $s$-wave scattering length for $^{86}$Sr atomic collisions \cite{skt10,mmp08}.
 

 


\section{Experimental Setup\label{Section:Experimental Setup}}
\subsection{Laser Cooling and Trapping}
Two-photon spectroscopy is performed on ultracold $^{86}$Sr atoms in a single-beam optical dipole trap (ODT) generated from a 1064-nm laser propagating perpendicular to gravity with beam waists of $260$\,$\mu$m and $26$\,$\mu$m \cite{mmp08,ssk14}. The tight waist provides vertical confinement. The trap depth after an evaporative cooling stage determines the sample temperature, which is set between $30-1000$\,nK. Typical atom numbers are several hundred thousand and peak densities are as high as $2\times 10^{12}$\, cm$^{-3}$. The number of atoms and sample temperature are measured using time-of-flight absorption imaging operating on the $^1S_0$-$^1P_1$ transition. Trap oscillation frequencies are determined by measuring dipole and breathing collective mode frequencies, which allow determination of trap volume and sample density.



\begin{figure}
  \includegraphics[width=3.3in,clip=true, trim=00 00 0 0, angle=0]{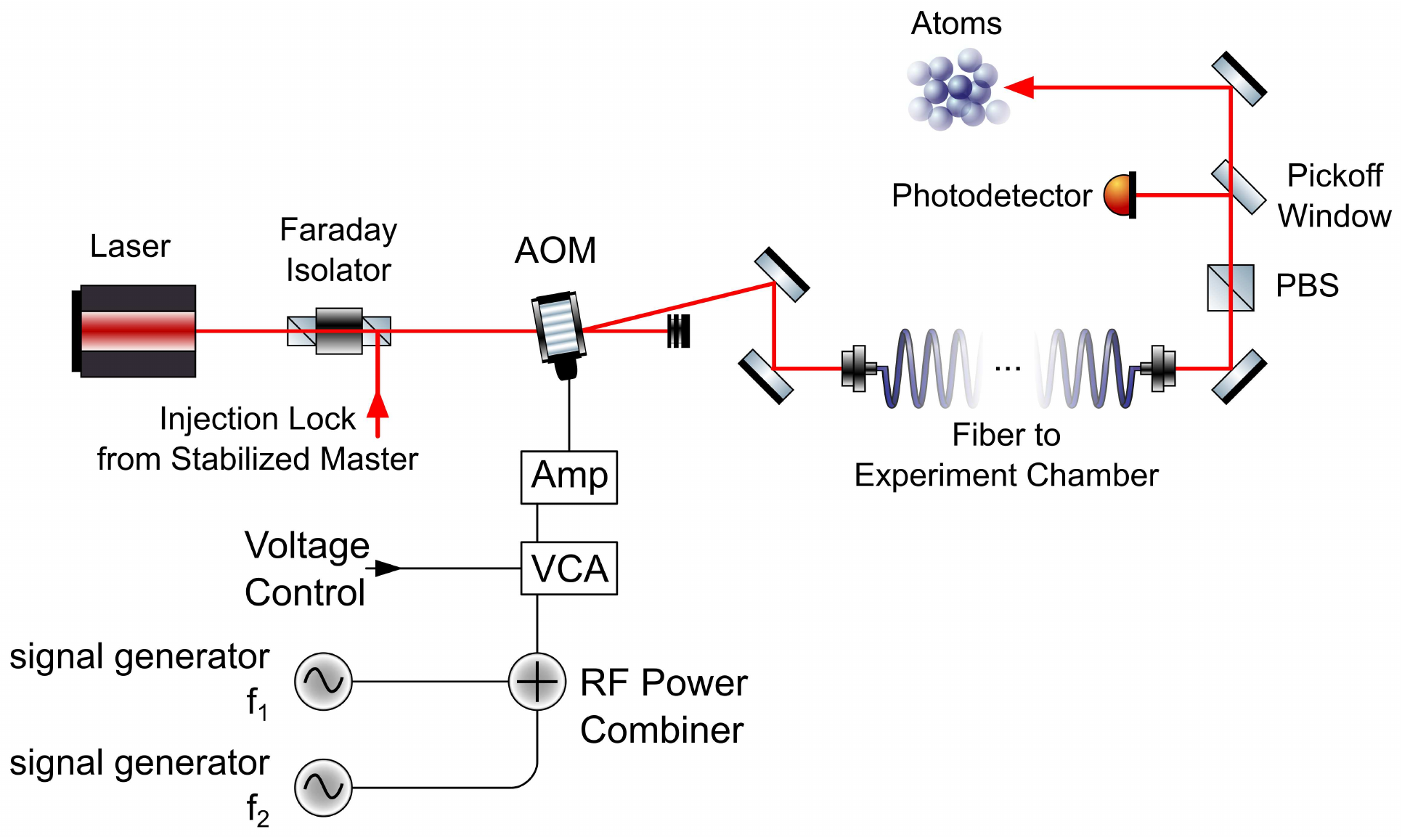}
  \caption{Photoassociation laser schematic (color online). A master laser is frequency-stabilized via saturated absorption spectroscopy to the $^1S_0$-$^3P_1$ atomic transition. After amplification with a diode slave laser, light at two controllable frequencies is generated with a single acousto-optic modulator (AOM) and delivered to the atoms with an optical fiber. The beat note between the two frequencies is monitored after the fiber.}
  
  \label{Experimental Setup}
\end{figure}

\subsection{Photoassociation}

After the atoms have equilibrated in the final ODT configuration, the PA lasers are applied (Fig.\ \ref{PASDiagram}). A single acousto-optic modulator, driven with two RF frequencies, is used to generate both PA beams. Light is derived from a frequency-stablized master laser (Fig.\ \ref{Experimental Setup}) and coupled into a single-mode optical fiber with output optics that yield a 320\,$\mu$m waist at the atoms, much larger than the size of the atom cloud. Both PA beams are linearly polarized along the same direction. The beat signal of the two light fields after the fiber is monitored on a photodiode and the RF powers are adjusted to ensure matched intensities for the two frequency components ($I_1=I_2\equiv I$).



The sample temperature is low enough that collisions are entirely $s$-wave. The target state for the two-photon transition has total angular momentum $J=0$ and binding energy $E_{b2}(<0)$. $^{86}$Sr has no nuclear spin and a $^1S_0$ electronic ground state, leading to a single ground electronic molecular potential (X$^1\Sigma_g^+$). The dominant intermediate state ($b_1$) is the $J=1$ rotational state of the second least-bound ($\nu=-2$) vibrational level on the $0^+_u$ molecular potential, which asymptotically connects to the $^1S_0$-$^3P_1$ atomic transition at long range. This state is bound by $44.246(10)$\,MHz \cite{bmc14}. We define $\Delta_1=\omega_1-E_{b1}/\hbar$ and $\Delta_2=\omega_1-\omega_2-E_{b2}/\hbar$ as the one-photon detuning from state $b_1$ and two-photon detuning from state $b_2$ respectively for an initial scattering state with collision energy $\epsilon=0$.
$\Omega_{2,12}$ is the Rabi frequency for coupling between states $b_1$ and $b_2$ due to the laser field at $\omega_2$ with single-beam intensity $I_2$.
Because the binding energy of the halo molecule is very small compared to $\Delta_1$, both laser frequencies are near resonance with the $\nu=-2$ state.  The transitions to the least-bound ($\nu=-1$) $J=1$ excited molecular state, bound by $1.633(1)$\,MHz, and the excited atomic state lie near enough in energy that they can effect our observations.



PA leads to loss of atoms from the trap through mechanisms that will be discussed below. The PA spectrum is obtained by holding $\omega_2$ fixed and varying $\omega_1$, which varies $\Delta_2$ across resonance (Fig.\ \ref{PASDiagram}). $\Delta_1$ thus also varies slightly during a scan, but the spectra are so narrow compared to $\Delta_1$ that we neglect this in our analysis. After an exposure time on the order of one hundred milliseconds, the number of ground-state atoms remaining and the sample temperature are measured with time-of-flight absorption imaging.


\section{Theoretical Description of Photoassociative Loss
\label{Theoretical Description of Photoassociative Loss}}

PA loss is described with a local equation for the evolution of the atomic density
\begin{equation}\label{densitydecay}
	\dot{n}=-2 Kn^2-\Gamma n,
\end{equation}
where the laser-frequency dependence of the collision-event rate constant, $K$, determines the spectrum of the PA loss. The one-body loss rate, $\Gamma$, is due to background collisions and off-resonant scattering from the PA lasers. By integrating this equation over the trap volume, we can obtain the evolution of the total number of trapped atoms
\begin{equation}\label{number}
   N(t)={N_0 \rm{e}^{-\Gamma t} \over 1+
   {2 N_0 \langle K \rangle V_2\over \Gamma V_1^2}(1-\rm{e}^{-\Gamma t})}
\end{equation}
where $N_0$ is the number of trapped atoms at the beginning of the PAS interaction time. The effective trap volumes $V_q$ are defined in App. \ref{sectionappendix} [Eq.~(\ref{eq:effectivevolumes})]. $\langle K \rangle$ is the trap-averaged collision event rate constant
\begin{eqnarray}\label{equationKeffective}
  \langle K \rangle&=& \frac{1}{V_{2}}\int_{\mathrm{V}} d^3r \,e^{-\frac{2U(\mathbf{r})}{k_{B}T}} \nonumber \\
  &&\times \frac{1}{h\,Q_{T}} \int_{0}^{\epsilon_{\text{max}}({\mathbf{r}})}d\epsilon \vert S\vert^2 \,e^{-\epsilon/k_{B}T},
\end{eqnarray}
which is itself a thermal average of the scattering probability for loss ($\vert S(\epsilon,\omega_1,\omega_2,...,\mathbf{r})\vert^2$) over the collision energy $\epsilon$, with an energy cutoff $\epsilon_{\text{max}}$ to be discussed momentarily. The trapping potential is given by $U(\mathbf{r})=mgz +h\chi_{1064,\text{g}}I_{1064}(\mathbf{r})-\tilde{U}_{\text{min}}$, where $mgz$ is the gravitational potential at height $z$, $I_{1064}(\vec{r})$ is the intensity of the trapping light, and $\chi_{1064,\text{g}}=11$\,Hz/(W/cm$^2$) \cite{YeKatori2008} is proportional to the polarizability of ground state atoms due to $1064$\,nm light. $\tilde{U}_{\text{min}}$ is subtracted to set the potential at the trap minimum to zero. The spatial integral is restricted to regions around the trapping local minimum with $U(\mathbf{r})$ less than the trap depth \cite{ycm11}. Downhill regions on the other side of the saddle point defining the trap depth are excluded. The laser intensity profile is measured independently, and the potential is found to be consistent with measured trap oscillation frequencies. The partition function is $Q_{T}=\left({2\pi k_{B}T \mu \over h^2}\right) ^{3/2}$ for reduced mass $\mu=m/2$ and sample temperature $T$, for atoms of mass $m$.

Equation (\ref{equationKeffective}) provides the correct thermal average when the collision-energy distribution does not need to be truncated $(\epsilon_{\text{max}}\rightarrow \infty)$. For our data, however, the ratio of sample temperature to trap depth is $k_BT/U_{\text{depth}}\approx 3$ for samples with temperature above $100$\,nK and drops to unity for 30\,nK samples, so truncation effects are important. If the single-particle kinetic-energy distribution function is a Boltzmann truncated at $U_{\text{depth}}-U(\mathbf{r})$, then the collision-energy distribution follows a Boltzmann distribution at low energies $(\epsilon\ll U_{\text{depth}}-U(\mathbf{r}))$ and falls off more quickly at larger energies, reaching zero at $2(U_{\text{depth}}-U(\mathbf{r}))$. We find that this treatment predicts a narrower distribution on the red side of the spectral line than we observe in our data, suggesting the presence of atoms in non-ergodic orbits with energies above the saddle point of the trap. This is not surprising given the large collisional loss rate associated with near-resonant scattering in this isotope. Fortunately, the molecular binding energy is strongly determined by the sharp edge of the spectrum on the blue side of the line, which is relatively insensitive to the description of the red tail. Our data is well fit with a truncated Boltzmann distribution of collision energies [Eq.~(\ref{equationKeffective})]. To estimate the systematic uncertainty introduced by this treatment, we perform fits with $\epsilon_{\text{max}}$ equal to $2(U_{\text{depth}}-U(\mathbf{r}))$ and $U_{\text{depth}}-U(\mathbf{r})$ and take the mean of the two results as the best value for the binding energy and half the difference as a systematic uncertainty $\sigma_{\epsilon_{\text{max}}}\approx 100$\,Hz. This procedure does not correctly represent the overall normalization of $\langle K \rangle$, but we are not concerned with overall signal amplitude in this study. Atom temperatures vary by no more than 20\% during the interaction time, so assuming a constant sample temperature is reasonable.



Bohn and Julienne \cite{bju96} provide an expression for $\vert S(\epsilon,\omega_1,\omega_2,...)\vert^2$ for a collision on the open channel of two ground-state atoms (g) with total energy $\epsilon$ leading to loss-producing decay from the excited state $b_1$ with rate $\gamma_1$. (See Fig.\ \ref{PASDiagram}.
More details are provided in App.\ \ref{sectionappendix}). This approach was found to be sufficient for describing two-photon spectroscopy to a more deeply bound molecular level in $^{88}$Sr \cite{mmp08}.

For the experiments reported here, we take the intermediate state $b_1$ as the $\nu=-2$ state and maintain significant intermediate-state detuning, $\Delta_1$, for which $|\Delta_1|\gg |\Omega_{2,12}|$. Thus we are in a Raman configuration, and not in the Autler-Townes regime \cite{mmp08}.  In the Raman regime, the expression from \cite{bju96} for the scattering probability $\vert S\vert^2$ for a given initial scattering energy will show a maximum near two-photon resonance at $\Delta_2+\epsilon/\hbar =\Omega_{2,12}^2/4\Delta_1$. Following a treatment discussed recently for a similar experiment in calcium \cite{pdt17}, if the detuning is restricted to near two-photon resonance then $\vert S\vert^2$ can be approximated as a Lorentzian [Eq.~(\ref{equationSprobLorentzian})] \footnote{Note that $\Omega_{2,12}$ is defined here to be the splitting of the Autler-Townes doublet one would observe in Autler-Townes configuration \cite{mmp08}, which differs from the Bohn-Julienne definition of the molecular Rabi coupling \cite{bju96}.}.





There are several concerns regarding the rigorous application of the Bohn and Julienne theory \cite{bju96} [App.\ \ref{sectionappendix}] to our experiment. The obvious one is that it assumes an isolated intermediate state, which is not always a good approximation because of the proximity of state $b_1$ to the $^1S_0+{^3P_1}$ asymptote and to the $\nu=-1$ state. Because of the small decay rate $\gamma_1$ of the intermediate molecular state associated with metastable $^3 P_1$ atomic state, we also expect that loss from the ground molecular state cannot be neglected.

The more subtle issue is that Eq.\ (\ref{equationSprobLorentzian}) is derived assuming only a single laser beam is near resonant with each leg of the two-photon transition, which is not a good approximation for two-photon spectroscopy of a halo state and the resulting small laser-frequency difference $\omega_1-\omega_2 \approx -E_{b2}\ll |\Delta_1|$. We can expect that coupling between pairs of states due to both photoassociation lasers will contribute to the transition strength and light shifts of the levels induced by the photassociation lasers \cite{bju96,bju99}.


In the absence of a more complete theory treating these effects, we analyze loss spectra using the effective expression
\begin{eqnarray}\label{equationApproxLorentzian}
  \vert S\vert^2 = \frac{\Gamma_L(\epsilon)+\gamma_{\text{eff}}}{\Gamma_L(\epsilon)}\hspace{1.5in} \nonumber \\
  \times \frac{\eta  A(\epsilon)} {\left(\omega_1-\omega_2+\epsilon/\hbar-E'_{b2}/\hbar\right)^2+\left[
  	\frac{\Gamma_L(\epsilon)+\gamma_{\text{eff}}}{2}\right]^2},
\end{eqnarray}
where the observed molecular binding energy ($E'_{b2}$) includes any perturbations due to AC Stark or collisional shifts, and
\begin{eqnarray}\label{ApproxLorentzianQuantitiesMain}
  A(\epsilon)&=& \frac{\Omega_{2,12}^{4}\gamma_1 \gamma_s(\epsilon)}{16(\Delta_1+\epsilon/\hbar)^4} \\  
  \label{ApproxLorentzianQuantities-2Main}
  \Gamma_L(\epsilon)&=& \frac{\Omega_{2,12}^{2}[\gamma_1 +\gamma_s(\epsilon)]}{4(\Delta_1+\epsilon/\hbar)^2}.
\end{eqnarray}
In practice, the variation of collision energy is negligible compared to the one-photon detuning $\Delta_1$. Also, ${\gamma}_{1}=2\gamma_{\text{atomic}}$, where $\gamma_{\text{atomic}}=4.7\times 10^4$\,s$^{-1}$ is the decay rate of the atomic $^3P_1$ level. ${\gamma}_{s}(\epsilon)$ is the stimulated width of $b_1$ due to coupling to the initial scattering state by laser 1, which for low energy can be expressed as \cite{ctj06,bmc14,pdt17}
\begin{equation}\label{equationstimulatedwidth}
	{\gamma}_{s}(\epsilon)=2k l_{\text{opt}} \gamma_1,
\end{equation}
where the optical length ($l_{\text{opt}}\propto I_1$) is related to the overlap between the initial colliding state and $b_1$, and $k=(2\mu \epsilon)^{1/2}/\hbar$. For the $\nu=-2$ intermediate state $l_{\text{opt}}/I=(1.5\pm0.3)\times 10^4\,a_0\mathrm{/(W/cm^2)}$ \cite{bmc14}, where $a_0=5.29\times 10^{-11}$\,m is the Bohr radius.

Two parameters have been added in Eq.\ (\ref{equationApproxLorentzian}) to account for deviations of the signal strength ($\eta$) and width ($\gamma_{\text{eff}}$) from the predictions of Eq.\ (\ref{equationSprobLorentzian}). If deviations from the single-channel theory of \cite{bju96} [Eq.\ (\ref{equationSprobLorentzian})] are small, we expect $\eta\sim1$, $\gamma_{\text{eff}}\sim 0$, and $E'_{b2}\sim E_{b2}+{\Omega_{2,12}^{2}}/{4(\Delta_1+\epsilon/\hbar)}$.


Light shifts (AC Stark shifts) due to the trapping lasers and collisions with ground-state atoms (density $n$) should contribute to shifts of molecular resonance. Similar effects were taken into account in a recent, high-precision study of weakly bound molecular states of ultracold ytterbium atoms \cite{bbc17}. In addition, we expect that both 689-nm excitation lasers will shift the line, not just $I_2\propto \Omega_{2,12}^{2}$. We model the relationship between the measured resonance positions and the unperturbed binding energy $E_{b2}$ as
\begin{equation}\label{Eq:GlobalFit}
	E'_{b2}=E_{b2}+h\chi_{689}I_{689}+h\chi_{1064}I_{1064}(\mathbf{r})+h\chi_{n}n(\mathbf{r}).
\end{equation}
The susceptibilities, in Hz per unit intensity or density, will be determined from experimental data or theoretical considerations. The variation with position of the trapping laser intensity ($I_{1064}$) and the density give rise to the spatial dependence of $|S|^2$ and the need for a spatial average in Eq.\ (\ref{equationKeffective}). We take $I_{689}$ as twice the single-beam intensity $I_{689}=2I$. The 689-nm excitation beam is large enough compared to the atom sample to neglect spatial variation. The functional form for the AC Stark shift due to the excitation lasers is discussed in Sec.\ \ref{sectionACStark}.



\begin{figure}
  \includegraphics[width=3.3in]{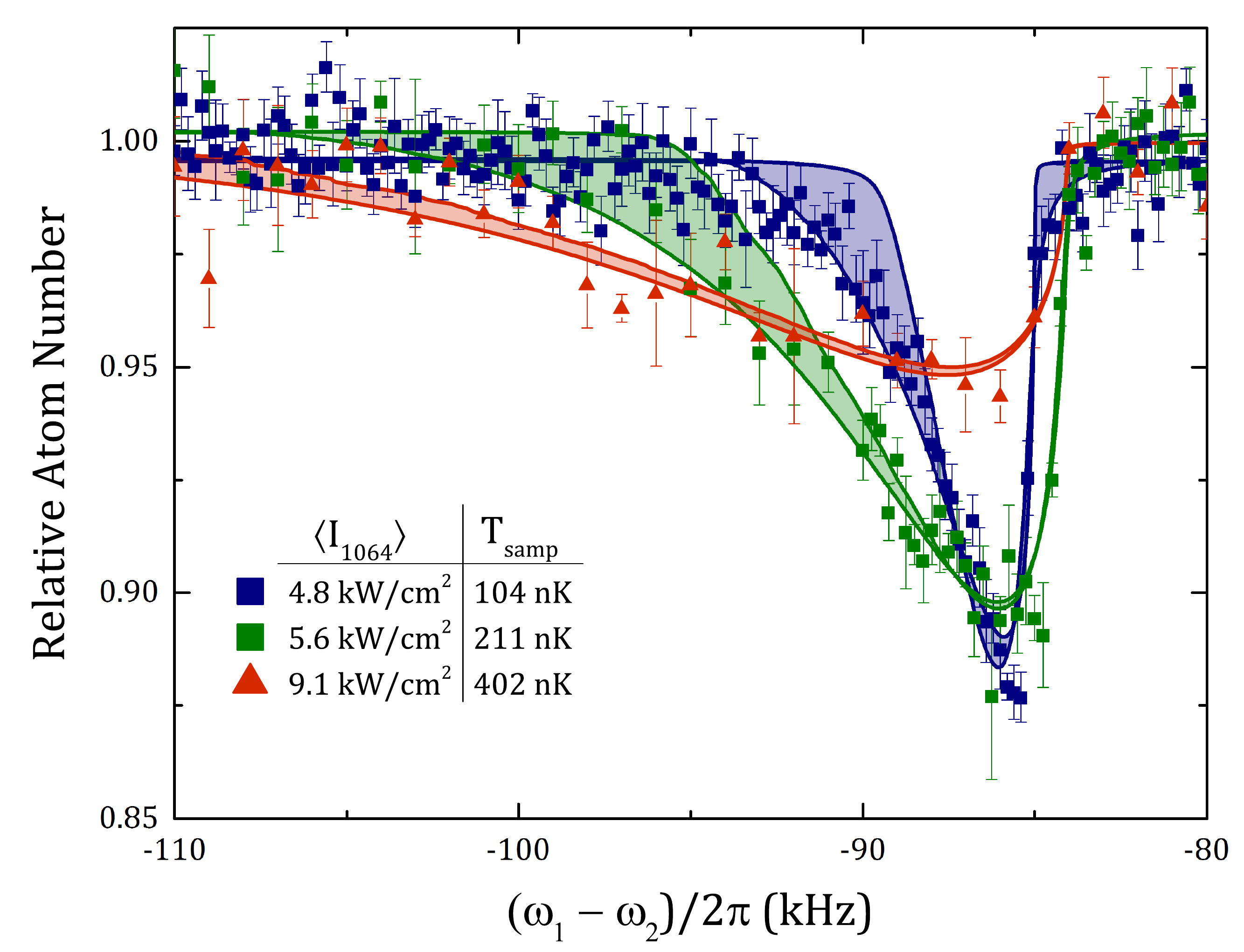}
  \caption{Atom-loss spectra as a function of two-photon difference frequency $(\omega_1-\omega_2)/2\pi$ for intermediate detuning
$\Delta_1/2\pi=-9$\,MHz. Sample temperature and average trapping laser intensity are indicated in the legend. The single-beam excitation laser intensity is $I=25$\,mW/cm$^{2}$ for the 104\,nK spectrum and $I=48$\,mW/cm$^{2}$ for the 211\,nK and 402\,nK spectra. Fits are described in the text, with the two boundaries of each band given by the fits with collision-energy truncation
$\epsilon_{\text{max}}$ equal to $2(U_{\text{depth}}-U(\mathbf{r}))$ and $U_{\text{depth}}-U(\mathbf{r})$.}
  \label{Fig:Spectraminus9MHzVaryTrapCold}
\end{figure}

\section{Spectral Fitting and Determination of the Halo Binding Energy}
\subsection{Fitting the Spectra}


Figure \ref{Fig:Spectraminus9MHzVaryTrapCold} shows a series of spectra for different final trap depths and sample temperatures. The characteristic asymmetric lineshape for excitation of a thermal sample is evident, with width decreasing as sample temperature decreases. The molecular binding energy is close to the sharp edge on the blue side of each spectrum.

We fit atom-loss spectra with Eq.\ (\ref{number}) for the evolution of atom number with time, using the phenomenological expression Eq.\ (\ref{equationApproxLorentzian}) for the scattering probability and Eq.\ (\ref{equationKeffective}) for the average of the collision event rate constant over the trap volume and collision energy. The sample temperature, perturbed resonance frequency $E'_{b2}$, $\eta$, and $\gamma_{\text{eff}}$ are taken as fit parameters. In the final analysis, temperatures are set to values determined from time-of-flight imaging of the atoms, but when they are allowed to vary, the fit values differ by no more than 10\%. Approximately 10 spectra are recorded for each set of experimental parameters, and the spread of resulting fit values are used to determine best values and uncertainties.

\subsection{AC Stark Shift due to Excitation Lasers}

\begin{figure}
  \includegraphics[width=3.3in,clip=true, trim=00 0 00 0, angle=0]{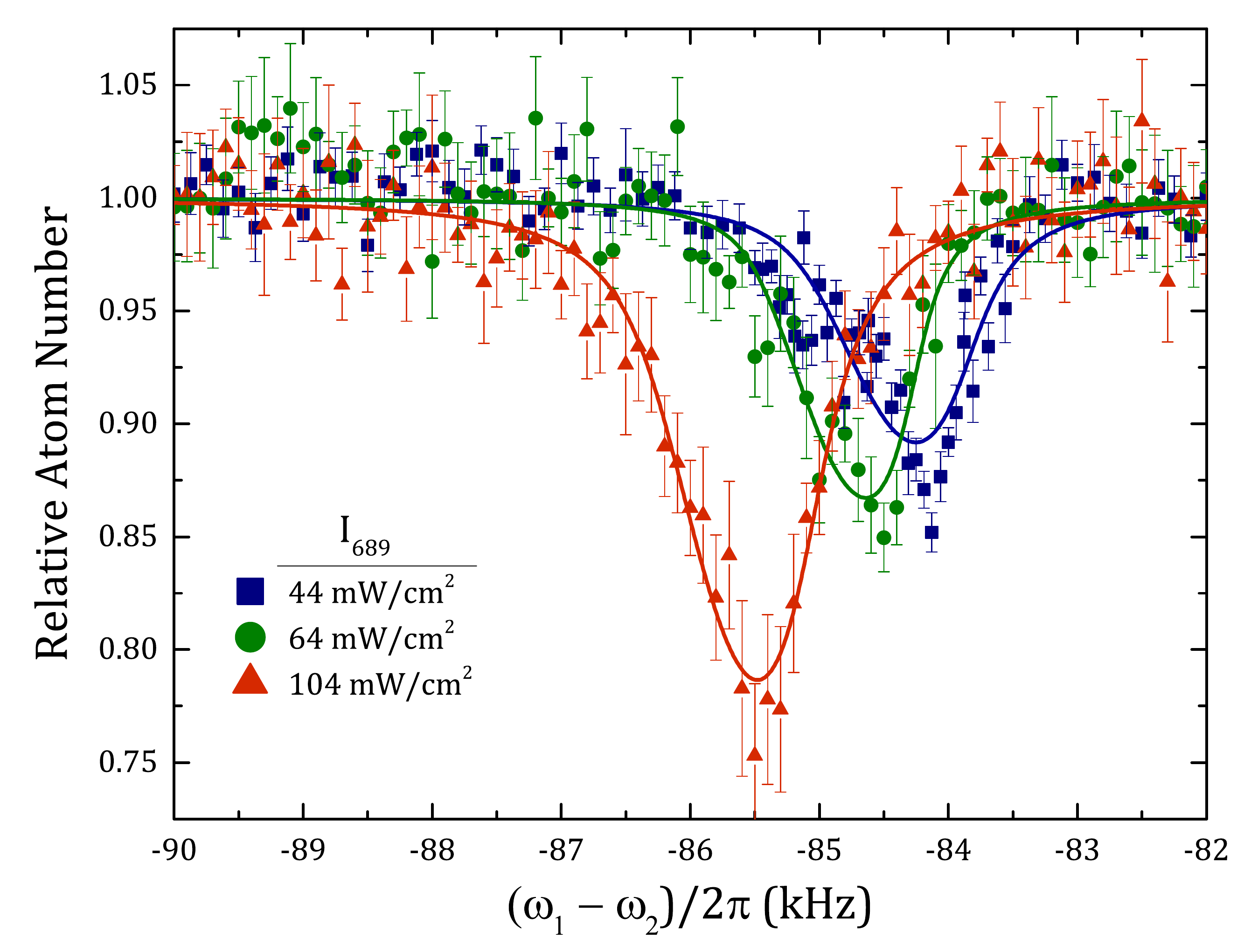}
  \caption{Atom-loss spectra as a function of two-photon difference frequency $(\omega_1-\omega_2)/2\pi$ for intermediate detuning $\Delta_1/2\pi=-9$\,MHz and various 689-nm excitation laser intensities. Twice the single-beam intensity $I_{689}=2I$ is indicated in the legend.}
  \label{Fig:SpectraVarying689Intensity}
\end{figure}

\begin{figure}
  \includegraphics[width=3.3in,clip=true, trim=00 0 00 0, angle=0]{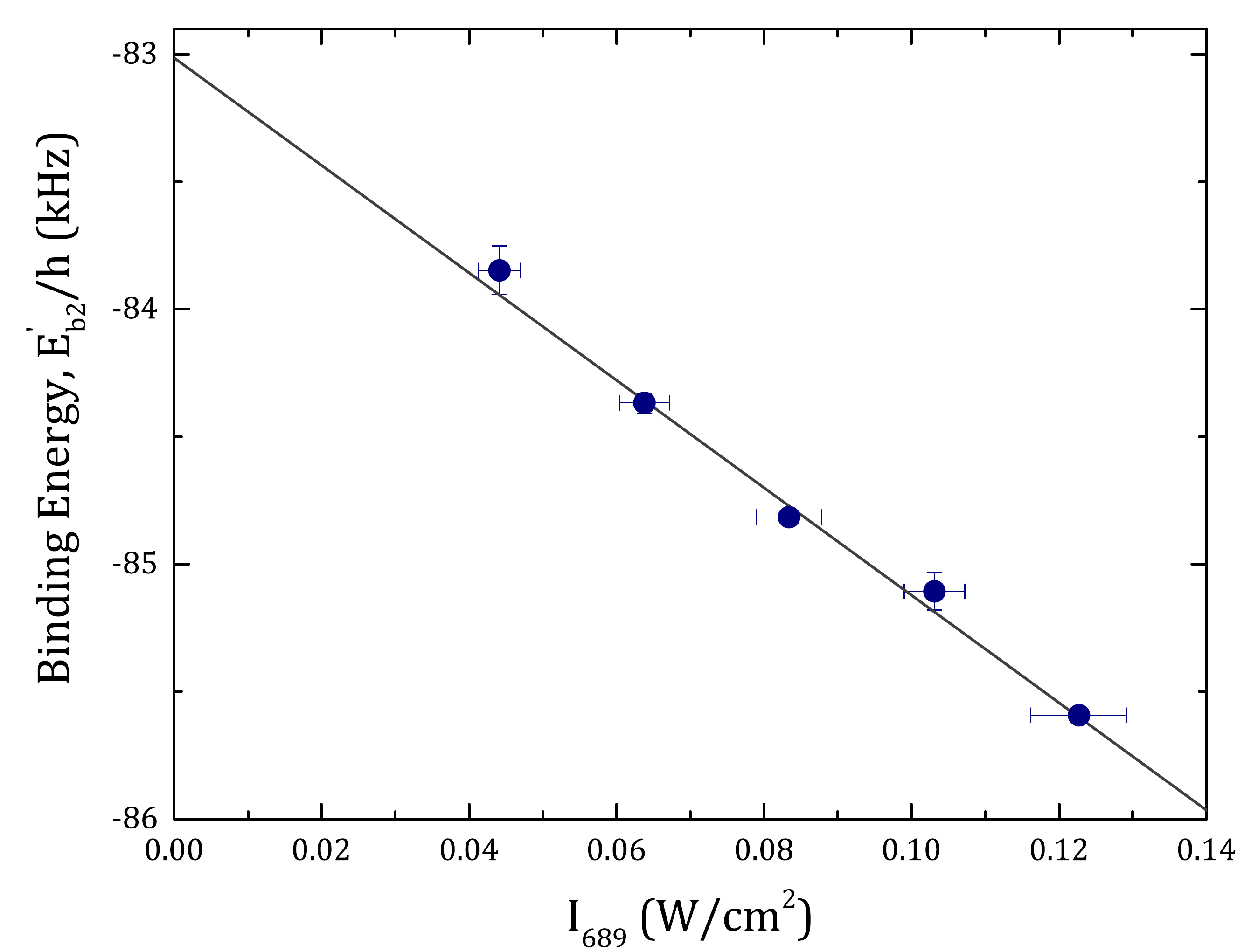}
  \caption{Measured resonance position $E_{b2}'$ plotted versus twice the single-beam intensity $I_{689}=2I$. The linear fit provides the AC Stark shift parameter $\chi_{689}$.}
  \label{Fig:ShiftWith689Intensity}
\end{figure}

The most significant perturbation to the resonance position is the AC Stark shift due to the excitation laser intensity, as shown in Fig.\ \ref{Fig:SpectraVarying689Intensity}. For this data, the trap parameters, temperature ($T=30$\,nK), and initial peak sample density ($n_0=2\times 10^{12}$\,cm$^{-3}$) are held constant. We vary the single-beam excitation intensity from $I=0.02-0.06$\,mW/cm$^{-2}$, and the excitation time is 50\,ms.
The observed shifts are comparable to the thermal width of the spectrum, allowing a precise determination of $\chi_{689}=-21(1)(2)$\,kHz/(W/cm$^{2})$ from a linear fit to the resonance positions, $E'_{b2}\propto h\chi_{689} I_{689}$ (Fig.\ \ref{Fig:ShiftWith689Intensity}). The first quoted uncertainty is statistical and it arises from variations in parameters and fluctuations in the measured intensity during the scans. The second value is systematic, reflecting uncertainty in laser-beam size and intensity profile at the atoms. Systematic uncertainty does not limit the accuracy of the extrapolation of the binding energy measurement to zero intensity. All parameters beside the 689-nm laser intensity are held fixed for this data set, and the AC Stark shift is not correlated with any other variable, such as density or trap intensity. We thus obtain an accurate measure of $\chi_{689}$ without attempting to account for other systematic shifts of $E'_{b2}$ in this data. A study of the dependence of $\chi_{689}$ on detuning from the excited molecular state
will be discussed in Sec.\ \ref{sectionACStark}.

Broadening to the red of the spectrum reflects the distribution of atom-atom collision energies, while broadening to the blue is most sensitive to decay of the intermediate state ($\Gamma_L$) and the phenomenological broadening term $\gamma_{\text{eff}}$ [Eqs.\ (\ref{ApproxLorentzianQuantities-2Main}) and (\ref{equationApproxLorentzian})]. The long lifetime of the excited state and the significant detuning $\Delta_1$ result in a width $\Gamma_L(\epsilon)$ less than 5\,Hz for all conditions. This is extremely small compared to observed width, which yields values of $\gamma_{\text{eff}}$ on the order of 100\,Hz. We hypothesize that this reflects decay of molecules in the electronic ground-state due to collisions with background atoms.

\subsection{Density-dependent Frequency Shift}
A shift of the two-photon resonance position is possible due to differing mean-field shifts of initial atomic and final molecular states arising from interaction with the background of ground-state atoms. Such a shift would be proportional to the atom density and depend upon the $s$-wave scattering lengths for atom-atom and atom-dimer collisions, $a_{86}$ and $a_{\text{ad}}$ respectively. This was observed in a Rb Bose-Einstein condensate (BEC) in \cite{wfh00}. For a non-degenerate gas, this effect yields $\chi_n=\hbar (\frac{a_{\text{ad}}}{\mu_{\text{ad}}}-4\frac{a_{86}}{\mu_{\text{aa}}})=\frac{\hbar}{m} (\frac{3 }{2}a_{\text{ad}}-8 a_{86})$, where $\mu_{\text{ad}}$ and $\mu_{\text{aa}}$ are the reduced masses for molecule-atom and atom-atom collisions respectively. Note that the shift would vanish for $a_{\text{ad}}=(16/3) a_{86}$.

The largest density used in our experiment ($\sim 1\times 10^{12}\,\mathrm{cm}^{-3}$) is relatively low compared to typical BEC densities, and at this time we are unable to accurately measure a variation of resonance position with density. However, the atom-atom scattering is close to resonance and thus Efimov physics can provide information on $a_{\text{ad}}$ \cite{bha07,nen17} and an estimate of the systematic error introduced by any residual density-dependent frequency shifts. For a zero-range interaction, the atom-dimer scattering length is related to the atom-atom scattering length through the three-body Efimov parameter $\kappa_*$ according to \cite{bha07}
\begin{equation}\label{Eq:EfimovMoleculAtomScatteringLength}
  a_{\text{ad}}=a_{86}\left\{1.46 + 2.15 \mathrm{cot}[s_0 \mathrm{ln} (14.1\kappa_* a_{86}) ]\right\}
\end{equation}
where $s_0=1.006$ \footnote{The Efimov parameter is related to $E^0_{3b}$ through $\kappa_*=(m|E^0_{3b}|/\hbar^2)^{1/2}$, where $E^0_{3b}$ is the binding energy the lowest Efimov trimer would have in the case of resonant atom-atom interactions.}.

In principle, the atom-dimer scattering length can take any value. However, for a deep atom-atom potential, such as for the ground-state strontium dimer \cite{skt10}, there is a universality of the three-body physics that sets $\kappa_*=0.226(2)/l_{\mathrm{vdW}}$ \cite{wie12}. Here, $l_{\mathrm{vdW}}=\left({2\mu C_6}/{\hbar^2}\right)^{1/4}/2=74.6$\,$a_0$ is the van der Waals length associated with the $C_6$ coefficient of the long-range Sr$_2$ ground-state potential. We use $C_6=3.03(1) \times 10^{-76}$\, J\,m$^6$ found from a fit of potential parameters to spectroscopic data \cite{skt10}, which is consistent with a recent \textit{ab initio} calculation \cite{zbb14}. This yields $\kappa_*=5.72\times 10^7$\,m$^{-1}=(330\,a_0)^{-1}$. Equation (\ref{Eq:EfimovMoleculAtomScatteringLength}) then predicts $a_{\text{ad}}=6.4\, a_{86}$, which leads to a small density-dependent frequency shift parameter of $\chi_n=50\,\mathrm{Hz}/(10^{12}\,\mathrm{cm}^{-3})$. A numerical calculation including a finite-range correction for the atom-atom interaction \cite{mwc17} results in $a_{\text{ad}}=3.5\, a_{86}$ and $\chi_n=-90\,\mathrm{Hz}/(10^{12}\,\mathrm{cm}^{-3})$. Thus, a very small shift is expected for the densities used here. We incorporate $\chi_n=0\pm 90 \,\mathrm{Hz}/(10^{12}\,\mathrm{cm}^{-3})$ as a set parameter in our model of the spectrum, where we set the systematic uncertainty to reflect the spread of theory predictions. This uncertainty will be significant for our determination of the unperturbed halo binding energy.

%

\subsection{Unperturbed Halo Binding Energy and AC Stark Shift due to Trapping Lasers}

\begin{figure}
 \includegraphics[width=3.3in,clip=true, trim=00 0 00 0, angle=0]{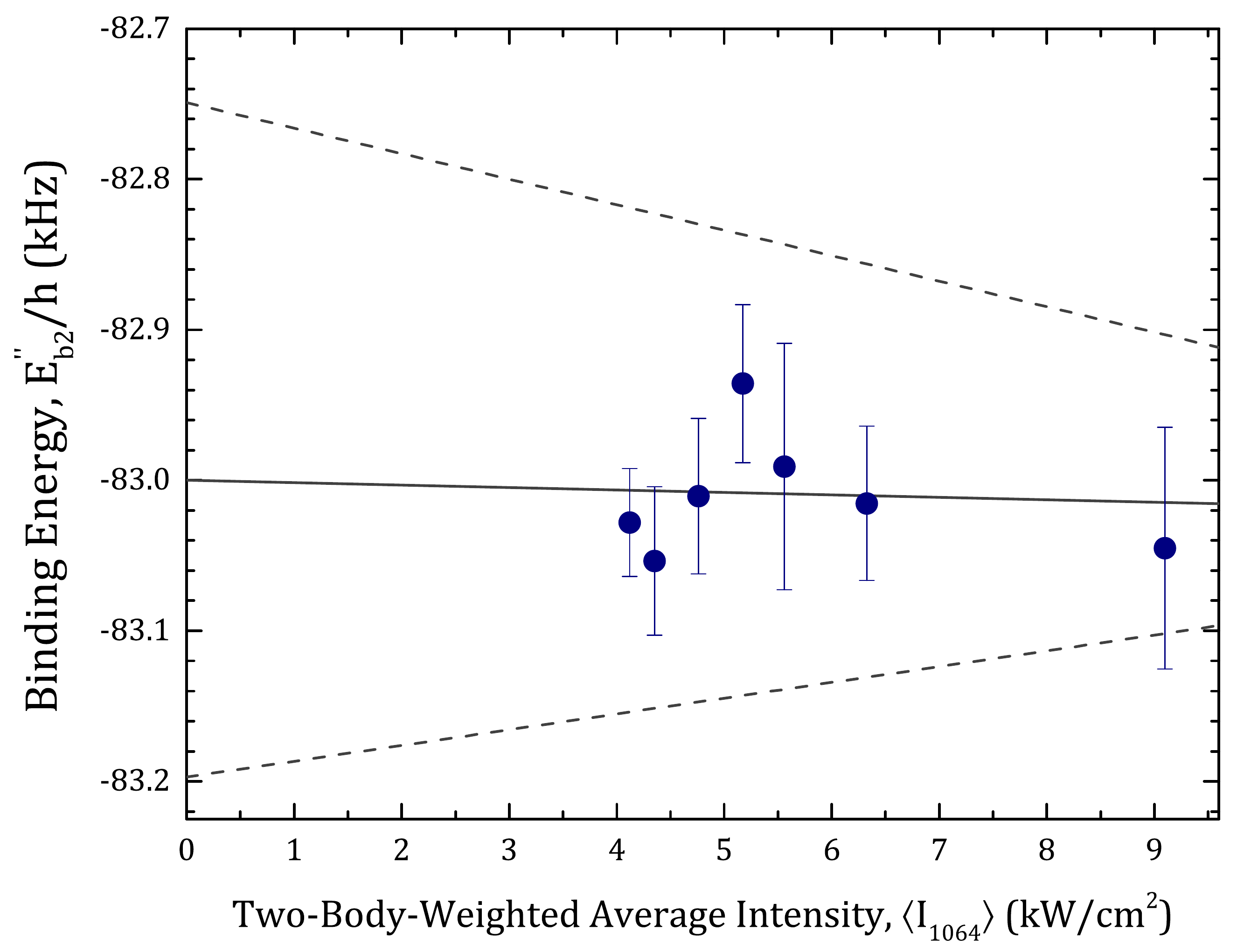}
  \caption{Measured resonance positions corrected for excitation-laser AC Stark shift and collisional frequency shift, $E_{b2}'-\chi_{689} I_{689} - \chi_{n}\langle n\rangle$, as a function of average trap laser intensity $\langle I_{1064} \rangle$ for the data such as in Fig. \ref{Fig:Spectraminus9MHzVaryTrapCold}. The trend line and confidence intervals are described in the text.
  }\label{Fig:ShiftWithTrapIntensity}
\end{figure}

With an accurate determination of $\chi_{689}$ and a value for $\chi_n$, we use the data shown in Fig. \ref{Fig:Spectraminus9MHzVaryTrapCold} to determine the susceptibility for the AC Stark shift from the trapping laser, $\chi_{1064}$, and the unperturbed halo binding energy $E_{b2}$. Figure \ref{Fig:ShiftWithTrapIntensity} shows a plot of $E_{b2}'-\chi_{689}I_{689} - \chi_{\text{n}}\langle n\rangle$ versus $\langle I_{1064} \rangle $, where $E_{b2}'$ is the resonance position from each fit and $\langle ... \rangle $ indicates a weighted average of the quantity over the trapped sample, with a weighting given by the square of atom density. This weighting reflects the contribution to photoassociative loss, a two-body process. The plotted uncertainties in $E_{b2}'-\chi_{689}I_{689} - \chi_{\text{n}}\langle n\rangle$ are from statistical variation in the fit parameters. The typical average density is $\langle n\rangle\approx 1\times 10^{12}$\,cm$^{-3}$. The linear fit function is to $E_{b2}+\chi_{1064}\langle I_{1064} \rangle $. In addition to statistical uncertainty, we have systematic uncertainty from $\chi_{\text{n}}$ and treatment of the truncation of the collision-energy integral [Eq.\ (\ref{equationKeffective})]. The dashed lines shown in Fig. \ref{Fig:ShiftWithTrapIntensity} are resulting fits when the values of $E_{b2}'-\chi_{689}I_{689} - \chi_{\text{n}}\langle n\rangle$ are shifted by the sum of these systematic uncertainties. The resulting value for the unperturbed binding energy is $E_{b2}/h=-83.00(7)(20)$\,kHz, where the first uncertainty is statistical, and the second is systematic. We observe a susceptibility to $I_{1064}$ of $\chi_{1064}=0\pm 10$\,Hz/(kW/cm$^2$).


\subsection{Discussion of the Halo Binding Energy}
\label{section:halostate}


In the limit of extremely small binding energy, and thus resonant atom-atom interactions, the binding energy of a halo molecule is approximately given by \cite{kgj06}
\begin{equation}\label{Eq:HaloEnergyNoCorrections}
	E_b=-\hbar^2/2\mu a^2.
\end{equation}
For interactions described at long-range by the van-der-Waals form, $V(r)=-C_6/r^6$, as with ultracold atoms, a convenient figure of merit for quantifying how accurate this simple expression should be is given by the ratio of the $s$-wave scattering length to the mean scattering length or interaction range, closely related to the van der Waals length through \cite{gfl93,cju05}
\begin{equation}\label{Eq:InteractionRangevdW}
  \bar{a}= l_{\mathrm{vdW}}\frac{\Gamma\left(\frac{3}{4}\right)}{\sqrt{2}\Gamma\left(\frac{5}{4}\right)}.
\end{equation}



Slightly away from resonance, corrections to the binding energy for the van der Waals potential were worked out in \cite{gao01,gao04}, yielding
\begin{equation}\label{Eq:BindingEnergyGao}
	E_{b2}=-\frac{\hbar^2}{2\mu(a-\bar{a})^2}\left[1+\frac{g_1\bar{a}}{a-\bar{a}}+\frac{g_2\bar{a}^2}{(a-\bar{a})^2} + ... \right],
\end{equation}
where $g_1=\Gamma(1/4)^4/6\pi^2-2=0.918...$ and $g_2=(5/4)g_1^2-2=-0.947...$. The range of validity of this expression extends to $a\sim 2 \bar{a}$. The accuracy of the first term in this expansion has been experimentally confirmed for various systems such as $^{85}$Rb \cite{ckt03,kgb03}, $^{40}$K \cite{rtb03,msg05} and $^{6}$Li \cite{bar05}. This derivation of Eq.\ (\ref{Eq:BindingEnergyGao}) assumes that the influence of short-range physics, which can be expressed through a quantum defect, varies negligibly from threshold to the molecular binding energy. We expect this to be an excellent approximation, since, as shown in Ref.~\cite{gao01} the corrections are typically less than about $1\%$ even for GHz binding energies.



\begin{figure}
  \includegraphics[width=3.3in,clip=true, trim=00 0 0 0, angle=0]{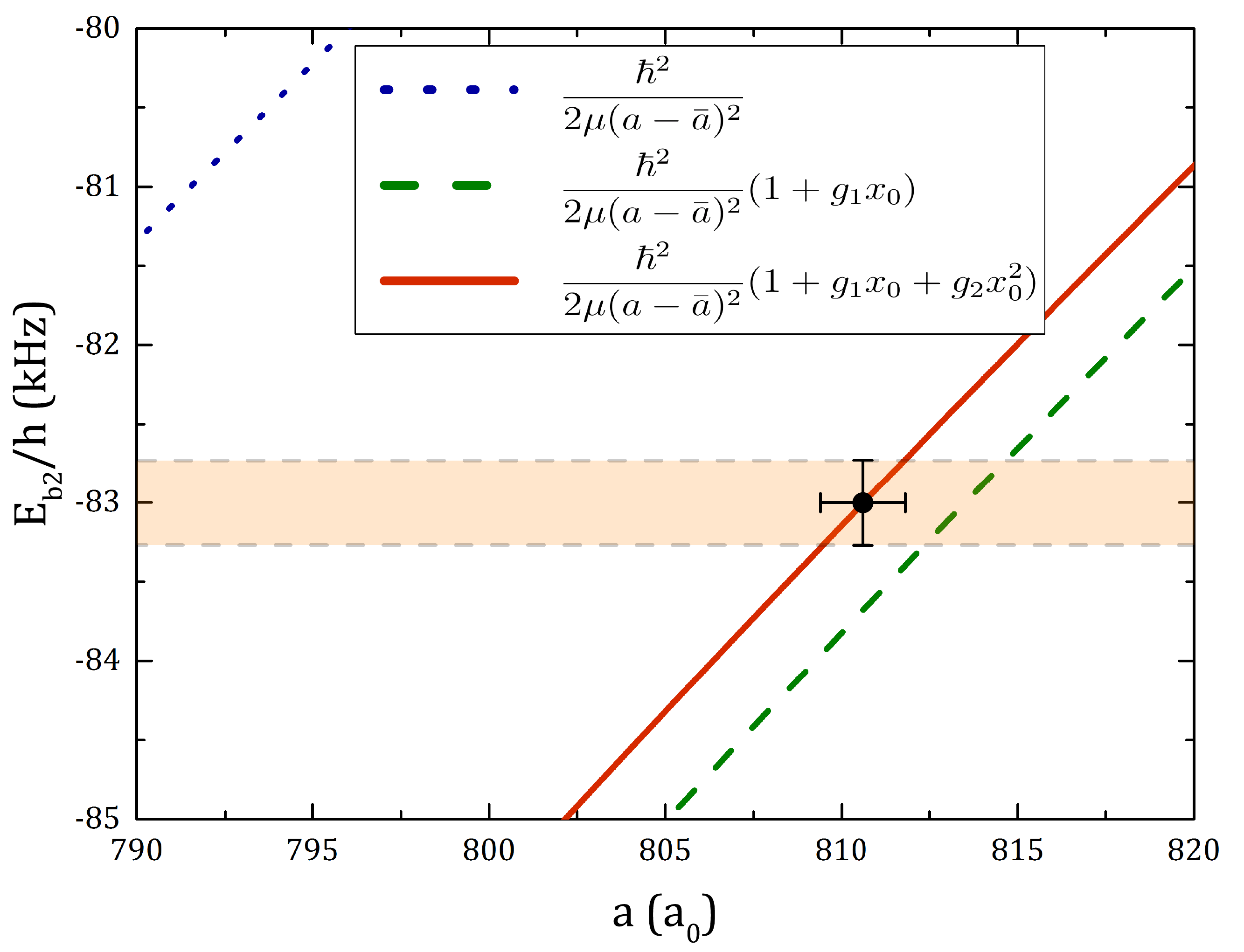}
  \caption{Halo binding energy versus $s$-wave atom-atom scattering length for $^{86}$Sr. The shaded region indicates our experimental measurement. The lines are predictions of Eq.\ \ref{Eq:BindingEnergyGao} retaining up to the first, second, and third terms as indicated in the legend [$x_0={\bar{a}}/({a-\bar{a}})$]. The data point is the prediction of Eq.\ (\ref{Eq:BindingEnergyGao}) for the recommended value of the measured binding energy.}
  \label{Fig:HaloBindingEnergy}
\end{figure}

For ground-state $^{86}$Sr atoms, $\bar{a}=71.3$\,$a_0$. The most accurate value available for the s-wave scattering length is $a=798 (12)$\,$a_0$ \cite{skt10}, satisfying the requirement of $a\gg \bar{a}$ for the least-bound state on the ground molecular potential to be a halo molecule. Nonetheless, ${\bar{a}}/({a-\bar{a}})=.10$, and the corrections given by Eq.\ (\ref{Eq:BindingEnergyGao}) are significant. Figure \ref{Fig:HaloBindingEnergy} shows the importance of the correction terms.

Equation (\ref{Eq:BindingEnergyGao}) and the previous best value of the scattering length \cite{skt10} predict a binding energy of $E_{b2}=-86(3)$\,kHz. This agrees with our measurement, but by inverting Eq.\ (\ref{Eq:BindingEnergyGao}), we can use our increased accuracy in $E_{b2}$ to extract an improved value of the scattering length of $a=810.6(12)$\,$a_0$, where uncertainty reflects the sum of the statistical and systematic uncertainty in $E_{b2}$. The next higher-order term in $x_0={\bar{a}}/({a-\bar{a}})$ is likely to introduce a correction on the order of $100$\,Hz in Eq.\ (\ref{Eq:BindingEnergyGao}), creating a systematic uncertainty in $a$ that is about one third of the uncertainty from our measurement.


\section{Functional Form and Frequency Dependence of AC Stark shift due to excitation lasers
\label{sectionACStark}}

\begin{figure}
  \includegraphics[width=3.3in,clip=true, trim=00 0 00 0, angle=0]{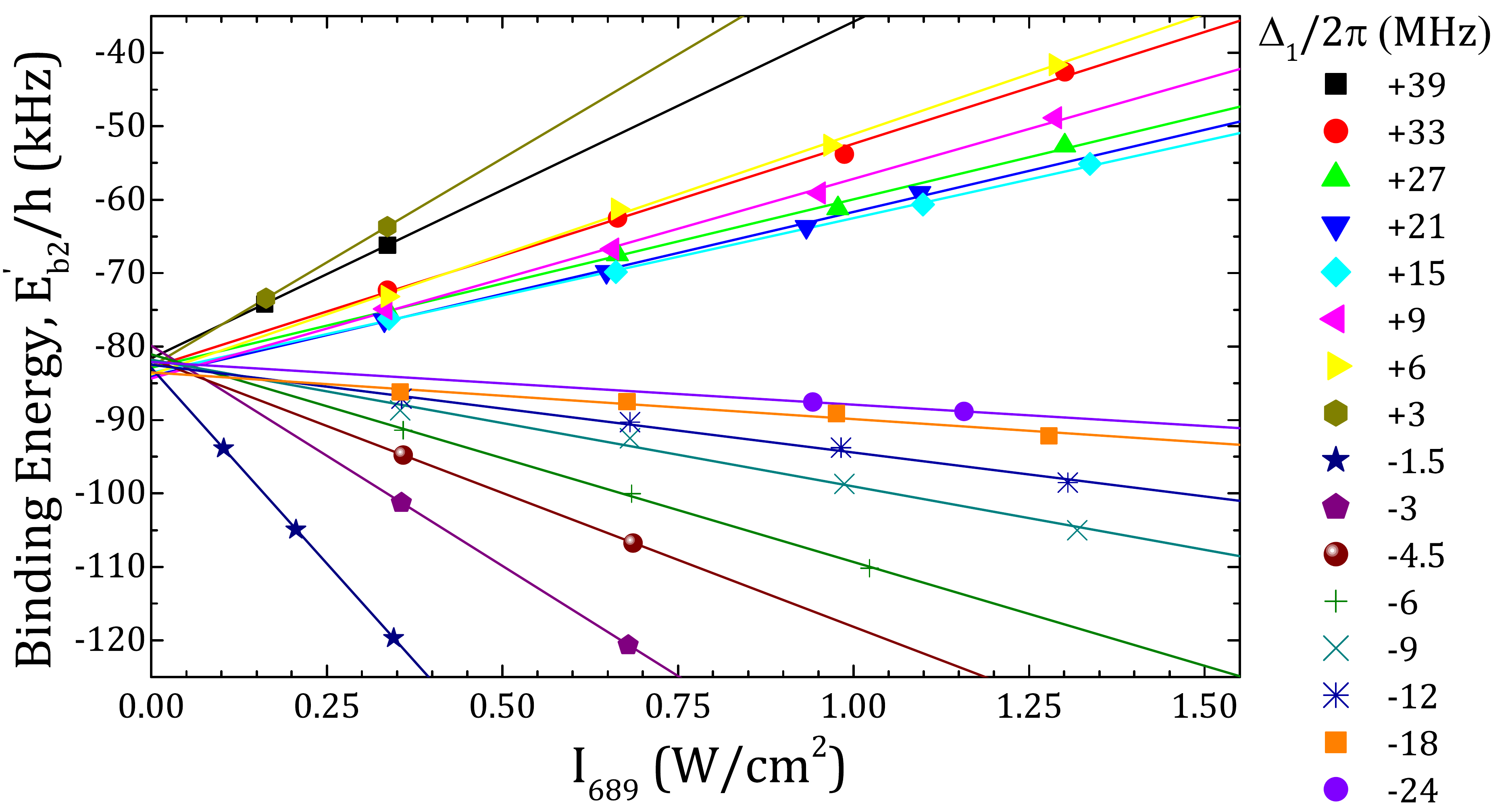}
  \caption{Two-photon PA resonance positions as a function of twice the single-beam excitation intensity, $2I=I_{689}$ for various intermediate state detunings, $\Delta_1$.
  }\label{Fig:ShiftsofLinesWithIntensity}
\end{figure}

The proximity of $^{86}$Sr to a scattering resonance and the susceptibility of the halo binding energy to the intensity of the excitation light suggests  using light to tune the binding energy and scattering length as was done with optically assisted magnetic Feshbach resonances \cite{blv09,chx15}. Understanding the frequency-dependence of $\chi_{689}$ is important for investigating this possibility, so we extracted this parameter from spectra at a wide range of 689-nm laser intensities and detuning from the intermediate resonance ($\Delta_1$). 

\begin{figure}
  \includegraphics[width=3.3in,clip=true, trim=00 0 0 0, angle=0]{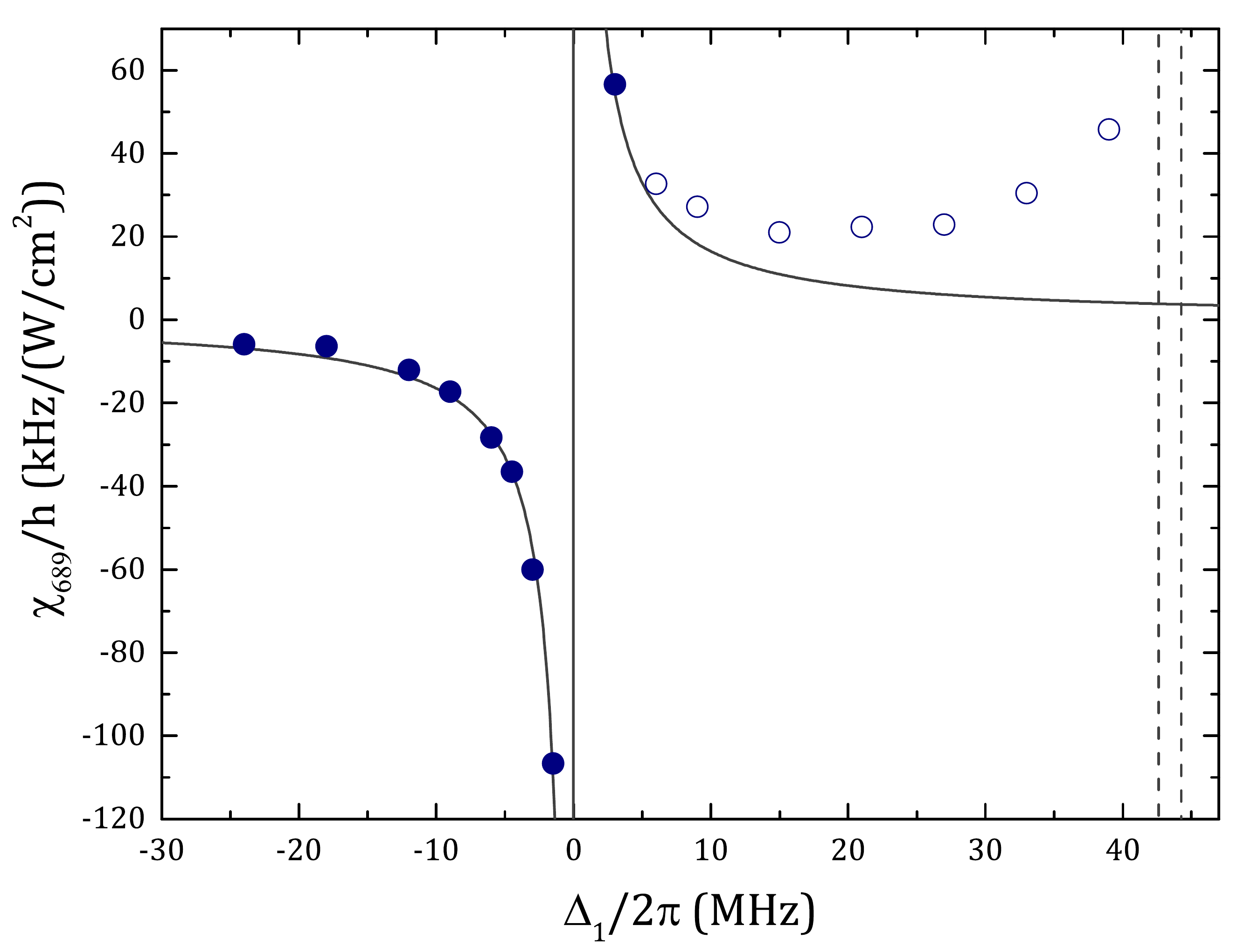}
  \caption{AC Stark shift susceptibility, $\chi_{689}$. Dashed lines indicate the positions of the $\nu=-1$, $J=1$ excited molecular state, bound by $1.633(1)$\,MHz, and the $^1S_0$+$^3P_1$ continuum. The solid line is a fit of $\chi_{689}\approx(\Omega_{12}/\sqrt{I})^2/(8\pi \Delta_1)$ to the solid circle data points.}
  \label{Fig:ACStark}
\end{figure}

Figure \ref{Fig:ShiftsofLinesWithIntensity} shows
the resulting resonance positions, $E'_{b2}$, versus twice the single-beam intensity, $2I=I_{689}$. The shift in molecular binding energy is linear with intensity over the explored range, but varies greatly in magnitude and sign. From linear fits, we extract the AC Stark shift parameter $\chi_{689}(\Delta_1)$ through $E'_{b2}\equiv E_{b2}+h\chi_{689}(\Delta_{1}) I_{689}$ (Fig.\ \ref{Fig:ACStark}).

In the experiment, the total 689-nm intensity oscillates with 100\% contrast according to $I_{\text{total}}=I_1+I_2+2\sqrt{I_1I_2}\cos \left[(\omega_1-\omega_2)t \right]=2I\left\{1+\cos \left[(\omega_1-\omega_2)t \right]\right\}$. The functional form we use to fit the AC Stark shift reflects the time average of the intensity and neglects the interference term. To confirm that this is the correct description, we numerically solved the time-evolution for a three-level system with similar optical couplings and oscillating optical intensity as present during two-photon PA of a halo state. The Hamiltonian is
\begin{eqnarray}\label{Eq:ThreeLevelHamiltonian}
H= \hspace{3in} \\
 \nonumber\\
\left(
    \begin{array}{ccc}
      0 & \Omega_{01}\left[\mathrm{cos}(\omega_1 t)+ \mathrm{cos}(\omega_2 t)\right] & 0 \\
      . & E_{b1} & \Omega_{12}\left[\mathrm{cos}(\omega_1 t)+ \mathrm{cos}(\omega_2 t)\right] \\
      . & . & E_{b2} \\
    \end{array}
  \right)
\nonumber
\end{eqnarray}
For $\Omega_{01}\ll \Omega_{12} \ll |\Delta_{1}|\equiv |\omega_1-E_{b1}/\hbar|$, which is analogous to the experimental conditions used here, we find that the two-photon resonance is shifted by
\begin{equation}\label{Eq:ACStarkFullModel}
\frac{\hbar\Omega_{12}^{2}}{4\Delta_{1}}+\frac{\hbar\Omega_{12}^{2}}{4\left(\Delta_{1}-E_{b2}/h\right)}\approx
\frac{\hbar\Omega_{12}^{2}}{2\Delta_{1}}.
\end{equation}
This agrees with our observation of a shift that is linear with intensity, and implies that the susceptibility is related to the Rabi frequency for a single-beam intensity $I$ through $\chi_{689}\approx(\Omega_{12}/\sqrt{I})^2/(8\pi \Delta_1)$.

This single-resonance model [Eq.\ (\ref{Eq:ACStarkFullModel})] describes the observed shifts well for detuning close to the $\nu=-2$ state of the $0^+_u$ molecular potential (small $\Delta_1$). For large positive $\Delta_1$, however, at which $\omega_1$ and $\omega_2$ approach atomic resonance, deviations indicate coupling to one or more other states (Fig.\ \ref{Fig:ACStark}). The most likely suspects are the $\nu=-1$, $J=1$ excited molecular state, bound by $1.633(1)$\,MHz, and the $^1S_0$+$^3P_1$ continuum. The sign of the deviation indicates that AC Stark shift of colliding $^1S_0$ atoms due to coupling to the $^3P_1$ state is dominant in this regime. We have neglected shifts due to collisions and the trapping laser, which are small at the large excitation-laser intensities used here.

%



A fit of the single-resonance model as shown in Fig.\ \ref{Fig:ACStark} yields $\Omega_{2,12}/2\pi\equiv\Omega_{12}/2\pi=800$\,kHz for $I=1$\,W/cm$^2$. 
Note that $\Omega_{2,12}$ as defined here would be the splitting of the Autler-Townes doublet \cite{pdt17}, which differs from the Bohn-Julienne definition of the molecular Rabi coupling \cite{bju96}. From the measured $\Omega_{2,12}$, one can extract the Franck-Condon factor, $f_{\text{FCF}}$, reflecting the overlap of the ground and intermediate molecular states through
\begin{equation}\label{Eq:FranckCondonRabiFrequency}
	\Omega_{2,12}=\sqrt{f_{\text{ROT}}}\sqrt{f_{\text{FCF}}}\gamma_{\text{atomic}}\sqrt{\frac{I}{2 _{\text{sat,atom}}}}
\end{equation}
where $I_{\text{sat,atom}}=2\pi^2\hbar c \gamma_{\text{atomic}}/(3\lambda^3)=3$\,$\mu$W/cm$^2$ is the atomic saturation intensity for the $^1S_0$-$^3P_1$ transition and $I=I_{689}/2$ is the single-beam intensity. The rotational factor $f_{\text{ROT}}$ accounts for the change in dipole moment from atom to molecule due to symmetry of the wave function and projection on a rotating molecular axis. Following the formalism described in \cite{pdt17}, $f_{\text{ROT}}=2$ for the $J=1\rightarrow 0$ bound-bound molecular transition studied here. This yields $f_{\text{FCF}}=0.03$.

\section{Conclusion\label{Conclusion}}

Using two-photon photoassociative spectroscopy, we have measured the binding energy of the least-bound vibrational level of the ground electronic state of the $^{86}$Sr$_2$ molecule. Using the universal prediction for the binding energy of a halo state including corrections derived for a van der Waals potential [Eq.\ (\ref{Eq:BindingEnergyGao})] \cite{gfl93,gao01,gao04}, we extract an improved value of the $s$-wave scattering length.

We also characterized the AC Stark shift of the halo-state binding energy due to light near resonant with the single-photon photoassociation transition. A model only accounting for a single excited-state channel \cite{bju96} cannot explain the observed frequency dependence of the AC Stark shift, which can be attributed to the proximity of other excited states.

Large AC Stark shifts of the halo state point to the possibility of optically tuning the $^{86}$Sr scattering length, similar to recent demonstrations of optical tuning of magnetic Feshbach resonances \cite{blv09,chx15}. This is attractive because ground-state strontium lacks magnetic Feshbach resonances. With improved measurement of the photoassociation resonance frequency and its dependence on background atom density, perhaps combined with optical manipulation of the scattering length, it may also be possible to study the landscape of Efimov trimers associated with this naturally occurring scattering resonance. This work also points to the need for improved theory, such as an improved calculation of the Sr ground-state molecular potential and $C_6$ coefficient, which could be compared with this high-accuracy measurement of the halo binding energy.




This work was supported by the Welch Foundation (C-1844 and C-1872) and the National Science Foundation (PHY-1607665). We thank Chris Greene for helpful discussions on Efimov physics.

\appendix

\section{Details of the Model of the Photoassociation Lineshape}
\label{sectionappendix}


PA loss is described with a local equation for the evolution of the atomic density [Eq.~(\ref{densitydecay})]. Integrating Eq.~(\ref{densitydecay}) over the trap volume yields the time evolution of the number of trapped atoms [Eq.~(\ref{number})]. The effective volumes used throughout this analysis are defined by
\begin{equation}\label{eq:effectivevolumes}
	V_{\text{q}}=\int_{\mathrm{V}} d^3r \, e^{-\frac{qU(\mathbf{r})}{k_{B}T}},
\end{equation}
for trapping potential $U(\mathbf{r})$. The collision event rate constant can be expressed as a thermal average of the scattering probability for loss, $\vert S(\epsilon,\omega_1,\omega_2,...,\mathbf{r})\vert^2$, over the collision energy $\epsilon$. We also average over the trap volume to allow for the possibility that the scattering probability can vary with position in the trap due to inhomogeneity of laser intensity profiles and the density distribution [Eq.~(\ref{equationKeffective})].


Bohn and Julienne \cite{bju96} provide an expression for $\vert S(\epsilon,\omega_1,\omega_2,...)\vert^2$ for a collision on the open channel of two ground state atoms (g) with total energy $\epsilon$ leading to loss-producing decay from the excited state $b_1$ with rate $\gamma_1$. (See Fig.\ \ref{PASDiagram}.) It yields
\begin{eqnarray}\label{equationSprob}
  \vert S\vert^2 =   \hspace{2.5in}&&\\
  {(\Delta_2+\epsilon/\hbar)^2{\gamma}_1{\gamma}_s \over
  	\left[(\Delta_1+\epsilon/\hbar)(\Delta_2+\epsilon/\hbar)-\frac{\Omega_{12}^{2}}{4}\right]^2+\left[ \frac{\gamma_1+\gamma_s}{2}\right]^2(\Delta_2+		 	\epsilon/\hbar)^2}, &&\nonumber
\end{eqnarray}
where all quantities are defined in the main text. For simplicity, we have omitted the light shift of $b_1$ due to coupling to the scattering continuum \cite{bju99}. Equation (\ref{equationSprob}) neglects all light shifts due to the trapping laser. Light shifts due to the photoassociation lasers coupling to states outside our model (Fig.\ \ref{PASDiagram}) are also neglected. The thermal energy is much greater than the zero-point energy for trap motion, $T\gg h\nu_{\text{trap}}/k_B$, so confinement effects are negligible \cite{zbl06}.




For the experiments reported here, we maintain significant intermediate-state detuning, $|\Delta_1|\gg |\Omega_{12}|$. Thus we are in a Raman configuration, and near two-photon resonance the expression for the scattering probability for a given initial scattering energy Eq.~(\ref{equationSprob}) can be approximated as a Lorentzian
\begin{eqnarray}\label{equationSprobLorentzian}
 \vert S\vert^2 \approx {A(\epsilon) \over
 \left(\Delta_2+\epsilon/\hbar-\frac{\Omega_{12}^{2}}{4(\Delta_1+\epsilon/\hbar)}\right)^2+\left[ {\Gamma_L(\epsilon)}/{2}\right]^2},
\end{eqnarray}
where $A$ and $\Gamma_L$ are defined in Eqs.\ (\ref{ApproxLorentzianQuantitiesMain}) and (\ref{ApproxLorentzianQuantities-2Main}).

As discussed in the text, we analyze loss spectra using the effective expression, Eq.\ (\ref{equationApproxLorentzian}) to account for possible deviations from the single-channel theory \cite{bju96}.

%


\end{document}